\documentclass[aps,prl,twocolumn,superscriptaddress]{revtex4-2}

\usepackage{graphicx}
\usepackage{amsmath}
\usepackage{amssymb}
\usepackage{hyperref}
\usepackage[utf8]{inputenc}
\usepackage{mathtools}
\usepackage[english]{babel}
\usepackage{bbm}
\hypersetup{colorlinks=true, linkcolor=blue, citecolor=blue, urlcolor=blue}
\usepackage{xcolor}
\usepackage{braket}
\usepackage[normalem]{ulem}

\makeatletter
\renewcommand{\fnum@figure}{FIG.~\thefigure}
\makeatother

\usepackage{bbold}

\usepackage{hyperref}

\begin{document}

\title{Nanoscale Mirrorless Superradiant Lasing}


\author{Anna Bychek}
\email{annabychek@gmail.com}
\address{Institut für Theoretische Physik, Universität Innsbruck, Technikerstr. 21a, A-6020 Innsbruck, Austria}

\author{Raphael Holzinger}
\affiliation{Department of Physics, Harvard University, Cambridge, Massachusetts 02138, USA}

\author{Helmut Ritsch}
\address{Institut für Theoretische Physik, Universität Innsbruck, Technikerstr. 21a, A-6020 Innsbruck, Austria}



\begin{abstract} 
We predict collective ’free-space’ lasing in a dense nanoscopic emitter arrangement where dipole-dipole coupled atomic emitters synchronize their emission and exhibit lasing behavior without the need for an optical resonator. At the example of a subwavelength-spaced linear emitter chain with varying fractions of pumped and unpumped emitters, we present a comprehensive study of this mirrorless lasing phenomenon. The total radiated power transitions from subradiant suppression under weak pumping to superradiant enhancement at stronger pumping, while exhibiting directional emission confined to a narrow spatial angle. At the same time multiple independent spectral emission lines below the lasing threshold merge towards a single narrow spectral line at high pump power. The most substantial enhancement and line narrowing occur when a fraction of unpumped atoms is present. We show that this leads to superradiant lasing near the bare atomic frequency, making the system a promising candidate for a minimalist active optical frequency reference.
\end{abstract}

\maketitle
%
\textit{Introduction.}---More than half a century after their first realization, lasers have proven to be a ubiquitous and versatile tool in wide areas of science and technology~\cite{chu1998nobel,hansch2006nobel,ludlow2015optical,haroche2025laser}. Generically, lasers share an optical resonator and a gain medium amplifying light via stimulated emission. Superradiant lasers are a special subclass of so-called single-mode lasers, where the spectral bandwidth of the optical gain medium is much smaller than the linewidth of the optical resonator modes. Here, the emission spectrum is locked around the gain maximum~\cite{norcia2016superradiance,Bohnet2012,zhang2012coherent,norcia2018frequency}. 
The most promising route to realizing a superradiant laser to date relies on a laser-cooled gas of individual atoms with a very narrow optical transition, often called a clock transition, providing for light amplification~\cite{ludlow2015optical}. Theoretical predictions suggest an unprecedented precision and accuracy of the emitted radiation at the bare atomic transition frequency with minimal sensitivity to technical noise~\cite{meiser2009prospects, reilly2025fully}, in close analogy to hydrogen maser technology. 
\begin{figure}[ht!]
    \centering
\includegraphics[width=0.95\columnwidth]{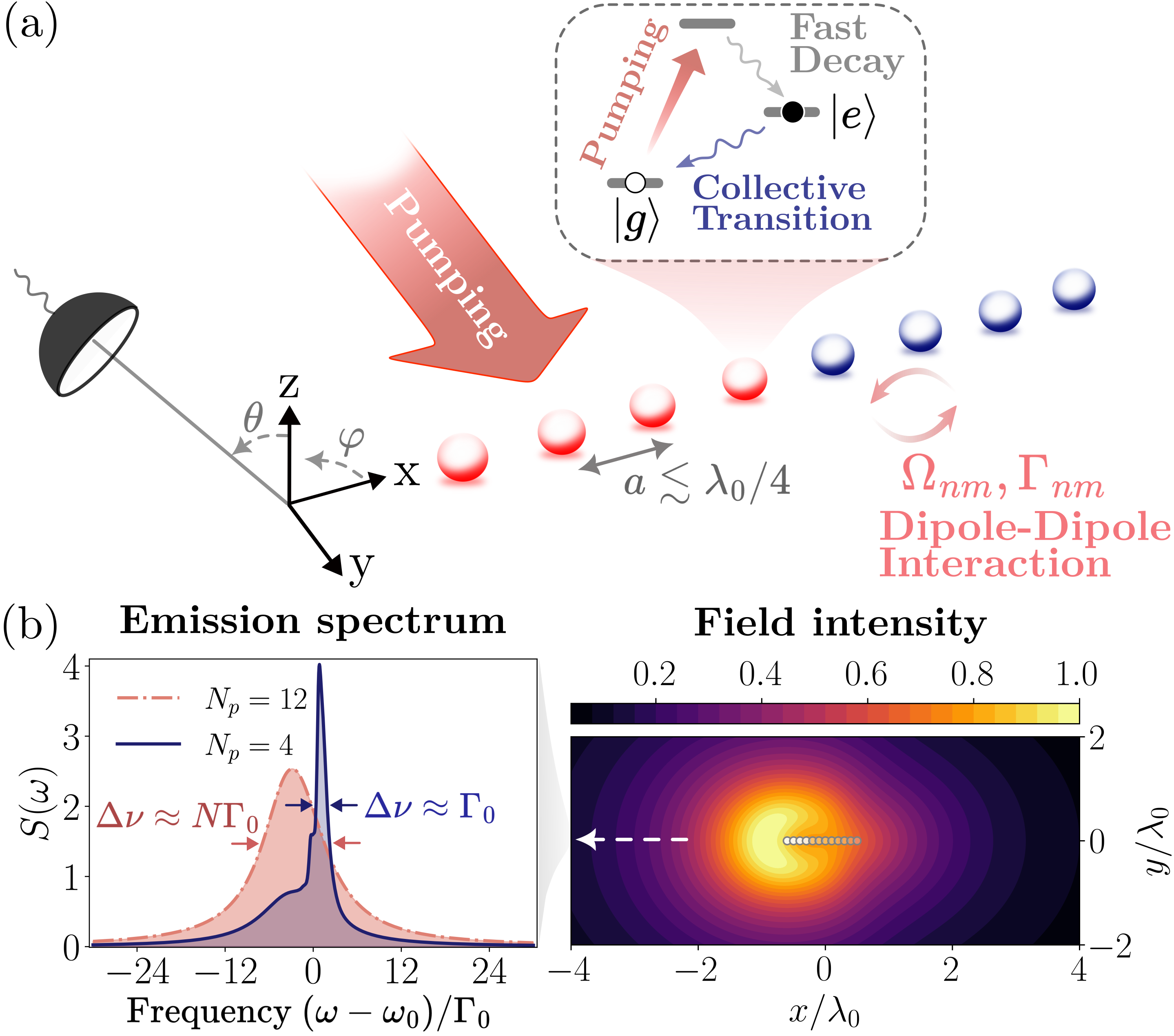}
    \caption{(a) Schematics: A linear chain of quantum emitters in free space with a spacing $a$ smaller than the $|g\rangle \leftrightarrow |e\rangle$ transition wavelength $\lambda_0$. The incoherent pumping of $N_p<N$ emitters in the chain creates a population in the excited state $|e\rangle$ from where the emitters decay at rate $\Gamma_0$ and interact collectively among each other through long-range dipole-dipole interactions. (b) Right panel: Directional steady-state superradiant emission from a chain of $N=12$ emitters with $N_p=4$ being pumped. The normalized electric field intensity distribution in the steady-state is plotted from the $z$-direction ($z/\lambda_0=1$) with the emitters linearly polarized perpendicular to the chain axis and spaced by $a/\lambda_0 = 0.1$. Left panel: The emission spectrum (Eq.~\eqref{spectrum}) in the direction of maximal emission $(\varphi,\theta)=(\pi,\pi/2)$ (dashed white line) exhibits a narrowing linewidth $\Delta \nu$ for partial pumping compared to a fully pumped ensemble. The incoherent pumping rate $R=10\Gamma_0$.}
    \label{fig1}
\end{figure}

Independent of these active clock developments over the years there have been several observations of narrow-band and directional coherent light emission from dense laser-driven atomic gases~\cite{lawandy1994laser,cao2000spatial,agarwal2024directional,Ferioli2023,ferioli2021laser}. While often there are nonlinear optical mixing processes or collective Raman scattering identified as the most likely origin of these phenomena, in some cases population inversion and lasing in driven ensembles appear to be the most plausible source of this radiation even without the presence of an optical resonator or external feedback~\cite{akulshin2018continuous, ramaswamy2023mirrorless,papoyan2019evidence,Scully1992}.~A detailed experimental observation and analysis of such a phenomenon were recently presented as an extremely bad-cavity limit in the context of a bad-cavity laser~\cite{zhang2024extremely}. There appears to be a connection to the phenomenon of random lasing eventually appearing in driven disordered media~\cite{wiersma2008physics,baudouin2013cold}~ or air lasing~\cite{li2019air}, which is sometimes tied to amplified spontaneous emission. 

For atomic dipole emitters localized at distances much smaller than the transition wavelength, common interaction with the radiation field results in collective super- and subradiance~\cite{han2025subwavelength,agarwal2024directional,Ferioli2023,ferioli2021laser,kusmierek2024emergence}. In the case of initial population inversion, one can observe superradiant emission bursts followed by subradiance periods with excitation energy storage much longer than the natural lifetime of the emitters~\cite{ferioli2021storage,asenjo2017exponential}. However, when the emitters are continuously repumped to their upper state and placed within an optical resonator with a large linewidth, superradiant stable steady-state emission of narrow-band radiation can occur, potentially creating an active optical frequency reference. We predict atomic dipole synchronization for dense nanoscopic emitter arrangements, exhibiting analogous behavior without any optical resonator.
As a motivating result to study mirrorless lasing in more detail, our earlier theoretical work showed that a ring-shaped nano-array of ground-state quantum emitters can serve as an optical resonator for a gain emitter placed in the center of the ring~\cite{holzinger2020nanoscale}, unexpectedly indicating a laser-like behavior for suitable operating conditions.

In this work, we investigate generic examples of regular subwavelength arrays of quantum emitters, where a certain fraction is incoherently driven to a non-thermal population inverted state. We focus on the spontaneous buildup of coherence between the individual quantum emitters, as well as the power, directional, and spectral properties of the collectively emitted radiation emerging from the nonlinear coupled dipole dynamics. Key questions are to identify the conditions for which continuous superradiance emerges in the system as well as its relation to conventional lasing. 

\textit{Theoretical model.}---We consider $N$ identical two-level emitters with spontaneous decay rate $\Gamma_0 = \omega_0^3 \mu_0^2 /(3\pi c^3 \epsilon_0 \hbar )$ tightly trapped in a regular spatial configuration at subwavelength distances and coupled via free space dipole-dipole interaction. Although we assume subwavelength distances between the emitters, they are still large enough so that we can neglect molecular interactions and use pairwise free space dipole-dipole interaction, including the radiative long-range contributions, which decrease as $1/r$ in the far field ($r$ being the separation between the emitters), leading to collective radiation properties via interference.

Our open system can thus be described by the following master equation in the Born-Markov approximation $\dot{\rho}= -i[\mathcal{H},\rho]+ \mathcal{L}_\mathrm{decay}[\rho] + \mathcal{L}_\mathrm{pump}[\rho]$~\cite{gardiner2004quantum,Carmichael:1993}, where $\hbar = 1$, $\rho$ is the emitter density matrix, and the Hamiltonian in the rotating frame of the emitter frequency $\omega_0$ reads
\begin{equation}
    \mathcal{H} = \sum_{n, m \neq n}^{N} \Omega_{nm} \sigma^+_n \sigma^-_m.
\end{equation}
The incoherent part describing the collective spontaneous emission is accounted for by the Lindblad term
\begin{equation}
    \mathcal{L}_\mathrm{decay}[\rho]=\frac{1}{2}\sum_{n,m}^N \Gamma_{nm} (2 \sigma^-_n \rho \sigma^+_m-\sigma^+_n \sigma^-_m \rho - \rho \sigma^+_n \sigma^-_m).
\end{equation}
The incoherent pumping process can be consistently modeled as inverse spontaneous emission, which populates the excited quantum emitter state without resonant coherent driving. Hence we add an extra Lindblad superoperator for the pump
\begin{equation} \label{pump}
    \mathcal{L}_\mathrm{pump}[\rho]=\frac{1}{2}\sum_{n=1}^N R_{n} (2 \sigma^+_n \rho \sigma^-_n-\sigma^-_n \sigma^+_n \rho - \rho \sigma^-_n \sigma^+_n),
\end{equation}
where the pump rate $R_n$ can vary as illustrated in Fig.~\ref{fig1}\hyperref[fig1]{(a)}, such that only a fraction of the emitters is pumped.

Dipole-dipole interaction is represented by the coherent and dissipative part
\begin{equation} \label{dipole-dipole}
\Omega_{nm} -\frac{i\Gamma_{nm}}{2} = -\mu_0 \omega_0^2  \mathbf{d}^*_n \cdot \mathbf{G}(\mathbf{r}_n-\mathbf{r}_m,\omega_0) \cdot \mathbf{d}_m 
\end{equation}
of the free space EM Green's function with $\mathbf{r}_n$ and $\mathbf{d}_n$ being  the position and dipole polarization of the $n^\mathrm{th}$ emitter, which reads~\cite{novotny2006principles}
\begin{align}
    \mathbf{G}(\mathbf{r},\omega_0) = \frac{e^{ik_0 r}}{4\pi k_0^2 r^3} &\bigg[(k_0^2 r^2 + ik_0 r -1) \mathbb{1}- \nonumber \\
    &-(k_0^2 r^2 + 3i k_0r - 3) \frac{\mathbf{r}\otimes \mathbf{r}}{r^2} \bigg],
\end{align}
where $r=|\mathbf{r}|$ and $k_0 = 2\pi/\lambda_0$ is the wavenumber corresponding to the emitter transition frequency. For a single emitter, Eq.~\eqref{dipole-dipole} yields the vacuum emission rate $\Gamma_{nn}=\Gamma_0$, while the energy shift $\Omega_{nn}$ arising from $\mathbf{G}$ function leads to a divergence and will be set to zero in the following, as it would lead to a re-normalized resonance frequency $\omega_0$. We will assume throughout this work that all emitters are linearly polarized in the $z$ direction $\mathbf{d} = (0,0,1)^T$. However, qualitatively similar conclusions are reached for other polarizations.

The total rate of the ensemble photon emission is then given by~\cite{allen_eberly_1975}
\begin{equation} \label{eq:emission}
    I(t)  = \sum_{n,m=1}^N \Gamma_{nm} \langle \sigma^+_n \sigma^-_m \rangle.
\end{equation}
Note that besides the individual diagonal terms in the final double sum, the relative phase coherence of emitter pairs in the off-diagonal terms can strongly modify the total emission power.
We will see later that these off-diagonal terms can be destructive (subradiance) or constructive (superradiance) depending on the chosen operating conditions and vanish in dilute emitter ensembles.
The electric field intensity $ \langle \mathbf{E}^-(\mathbf{r})  \mathbf{E}^+(\mathbf{r})\rangle $ at any position $\mathbf{r}$ can be directly calculated from electric field generated by the dipole operators via
\begin{equation} \label{e-field}
    \mathbf{E}^+(\mathbf{r}) = \mu_0 \omega_0^2 \sum_{n=1}^N \mathbf{G}(\mathbf{r}-\mathbf{r}_n,\omega_0) \cdot \mathbf{d}_n \sigma_n^-,
\end{equation}
where we assume the initial field to be in the vacuum state, neglecting any input noise~\cite{Carmichael:1993,gardiner2004quantum}.

To analyze the steady-state spectral properties of the system and particularly the linewidth $\Delta \nu$ of the emitted field, we calculate the directional spectrum $S(\omega,\theta,\varphi)$ via the Fourier transform of the electric field intensity in the far field ($r\gg \lambda_0$). Here, the angles $(\theta,\varphi)$ parametrize the direction of detection, shown in Fig.~\ref{fig1}\hyperref[fig1]{(a)}.
The directional far-field spectrum for an ensemble of atoms with identical dipole polarizations $\mathbf{d}$ and positioned along the $x$-direction is then given by~\cite{scully2006directed,allen_eberly_1975,Clemens2004directed,clemens2003collective,carmichael2000quantum,masson2020many}
\begin{widetext}
\begin{equation}
\label{spectrum}
S(\omega, \varphi,\theta) = \frac{3\Gamma_0}{8\pi} \Big(1- (\mathbf{R}(\theta,\varphi)\cdot \mathbf{d})^2\Big) \times 2 \Re \left\{ \int_{0}^{\infty} d\tau e^{-i\omega \tau} \sum_{n,m=1}^N  e^{i k_0 (x_m - x_n)\cos{\varphi}\sin{\theta}} \langle  \sigma^+_n(\tau) \sigma^-_m(0) \rangle \right\},
\end{equation}
\end{widetext}
with $\sigma^-_m(0)$ referring to the stationary state of $\sigma^-_m$ under the time evolution, and the unit vector $\mathbf{R}(\theta,\varphi)=[\cos(\varphi)\sin(\theta),\sin(\varphi)\sin(\theta),\cos(\theta) ]$ describing the direction of detection, see the Supplemental Material for details~\cite{supplemental}. In Fig.~\ref{fig1}\hyperref[fig1]{(b)}, we show the steady-state emission properties for a chain of $N=12$ emitters, where a fraction of $N_p$ emitters is incoherently pumped with the rate $R=10\Gamma_0$. The dipole moment orientation is chosen perpendicular to the chain direction, and the emitter spacing is $a/\lambda_0=0.1$. The electric field intensity distribution in the stationary state and the emission spectrum in the direction of maximum emission $(\varphi,\theta)=(\pi,\pi/2)$ are plotted. We observe the emergence of a narrow spectral peak close to the bare atomic transition frequency $\omega_0$ when only a fraction (in this case $N_p=4$) emitters are pumped compared to all of them being pumped. We will study the superradiance and spectral properties in more detail in the following sections.

\begin{figure}[t!]
    \centering
    \includegraphics[width=0.94\linewidth]{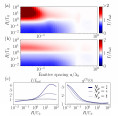}
    \caption{Collective enhancement of photon emission: Steady-state emission intensity $I/I_{ind}$ normalized to the independent emitter case ($a \gg \lambda_0$) as a function of the emitter spacing $a$ and pumping rate $R$, where in (a) the first $N_p=4$ emitters are pumped, whereas in (b) $N_p=N=8$ all emitters are pumped. The white region indicates the crossover from the steady-state subradiant (blue region) to superradiant emission (red region). (c) Parameter cut as a function of the pumping rate for $a=0.05\lambda_0$ for various pumped fractions as well as the second-order correlation function. Collective superradiant emission is enhanced in the presence of unpumped emitters due to dipole-dipole exchange interaction. Simulations are based on the full quantum master equation.}
    \label{fig:emission_rate}
\end{figure}

\textit{Gain and superradiance of a partially pumped array.}---In contrast to the collectively pumped array studied here, in a conventional laser the optical resonator and the gain medium are separate physical entities. In this section, we will use this analogy dividing our emitter ensemble into an antenna (resonator) and a gain subgroup, where only the latter is incoherently pumped. At the most basic example of a chain of $N$ quantum emitters with the first $N_p < N$ being pumped, we will demonstrate that this can lead to superior power emission at diminishing frequency shifts as desired for an active clock.

Fig.~\ref{fig:emission_rate} shows how for $N$ emitters the total photon emission in Eq.~\eqref{eq:emission} transitions from subradiance to superradiance compared to the case of independent emitters. The steady-state emission of $N_p$ independent emitters with pumping rate $R$ can be readily obtained and reads, $I_\mathrm{ind} = N_p\Gamma_0 R /(R+\Gamma_0)$. One can see the onset of superradiant emission starting from $a \approx 0.5 \lambda_0$, which becomes appreciable at $a \lesssim 0.25\lambda_0$
 as the coherent dipole-dipole interaction increases. 
 To evaluate the statistical properties of the emitted light one typically calculates the second-order correlation function at zero time delay. In~Fig.~\ref{fig:emission_rate}\hyperref[fig:emission_rate]{(c)}, we observe that $g^{(2)}(0)$ steady-state values are close to unity above the pumping threshold, in particular at spacings $a\lesssim 0.1\lambda_0$, where the emission is superradiant, see the Supplemental Material for details on the second-order correlation function calculations~\cite{supplemental}.

Furthermore, Fig.~\ref{fig:emission_rate}\hyperref[fig:emission_rate]{(a)} and~\hyperref[fig:emission_rate]{(c)} show that a lower fraction $N_p/N<1$ of pumped emitters lead to a larger collective gain of the total emitted power. This enhancement compared to independent emitters follows from the presence of coherent dipole-dipole exchange between all $N$ emitters in the ensemble, enabling excitation transfer to the unpumped emitters. Since this unpumped emitter fraction is not broadened by the incoherent pumping rate or power broadening, it acts as an effective high-Q resonator, which induces enhanced narrowband emission analogous to a Purcell enhancement. The pumped fraction then creates a broadband gain filtered and amplified by the unpumped emitters.

\begin{figure}[t!]
    \centering
    \includegraphics[width=0.97\linewidth]{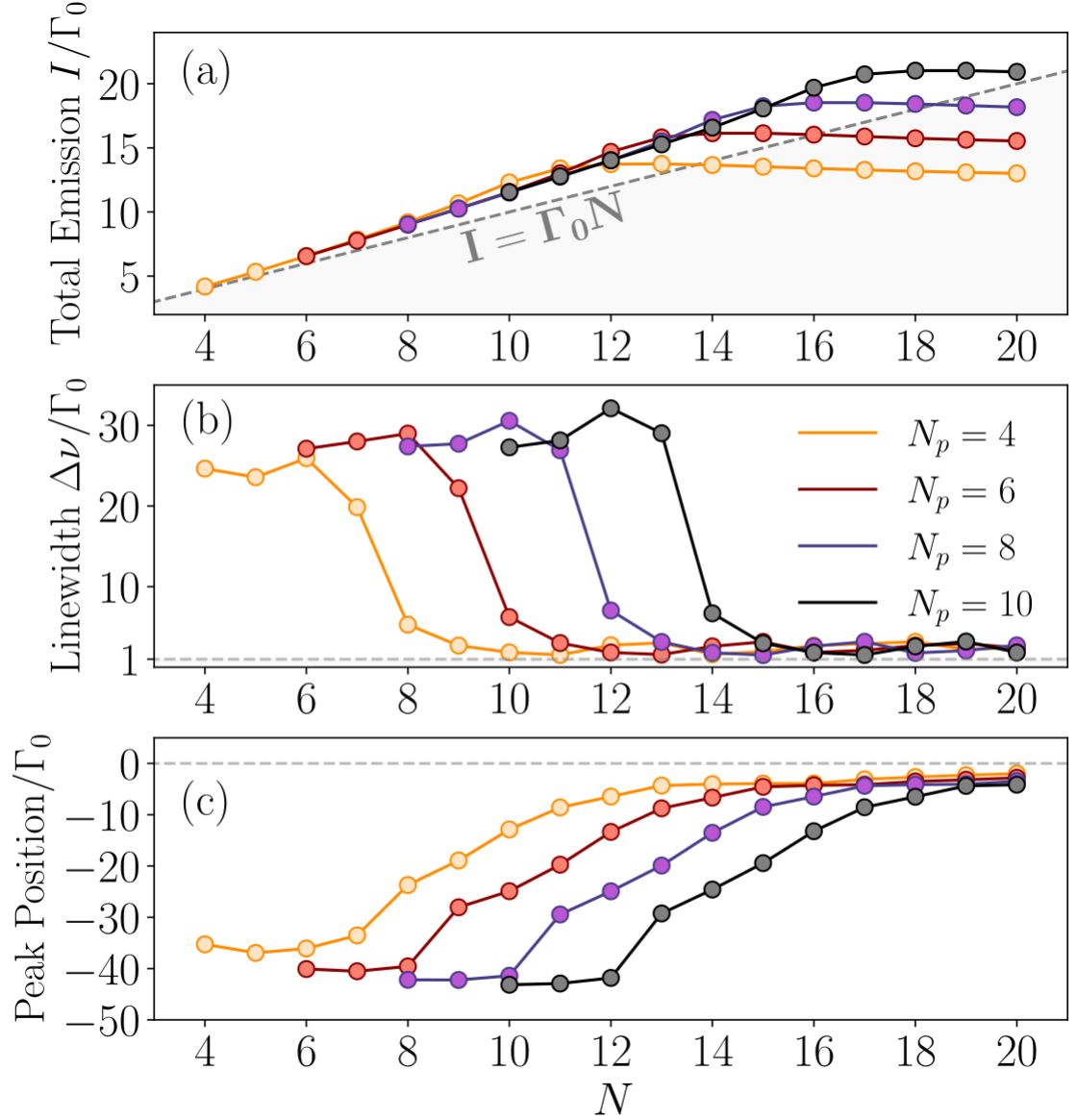}
    \caption{Narrow linewidth superradiance: Radiation from a linear chain of emitters spaced by $a=0.05\lambda_0$ with increasing total emitter number $N$ for a fixed number of pumped emitters $N_p$. (a) The steady-state emission power (Eq.~\eqref{eq:emission}) shows superradiance ($>N\Gamma_0$) and reaches saturation for large $N$. (b) The spectral linewidth $\Delta \nu$, evaluated in the direction of maximal emission (Eq.~\eqref{spectrum}) exhibits a sudden narrowing with increasing $N-N_p$ when $N$ reaches $N/N_p \approx 3/2$. (c) The spectral peak position $\omega-\omega_0$ approaches the atomic resonance even in the presence of strong dipole-dipole frequency shifts. In all plots: $R=20\Gamma_0$.}
    \label{fig3}
\end{figure}
 
\textit{Emission power and spectral linewidth using cumulant expansion.}---When one goes to larger systems beyond ten emitters, a numerically exact, full quantum simulation of the master equation gets soon out of reach of numerical capabilities. As we are centrally interested in gain and inversion, a common cutoff method towards the few excitation manifolds can not be used. Typically, a simple mean-field approach will not work as well, as no absolute phase will build up in the system, thus resulting in zero mean-field averages.
In the following, we will study the system dynamics using a second-order cumulant expansion~\cite{kubo1962generalized,plankensteiner2021quantumcumulantsjl,supplemental} including all pair correlations. Notably, in related physical models, this approximation turned out to capture the dynamics and steady-state properties remarkably well. For smaller systems, we benchmark the results against full quantum solutions of the master equation, see the Supplemental Material~\cite{supplemental}.

\begin{figure}[t!]
    \centering
    \includegraphics[width=1\columnwidth]{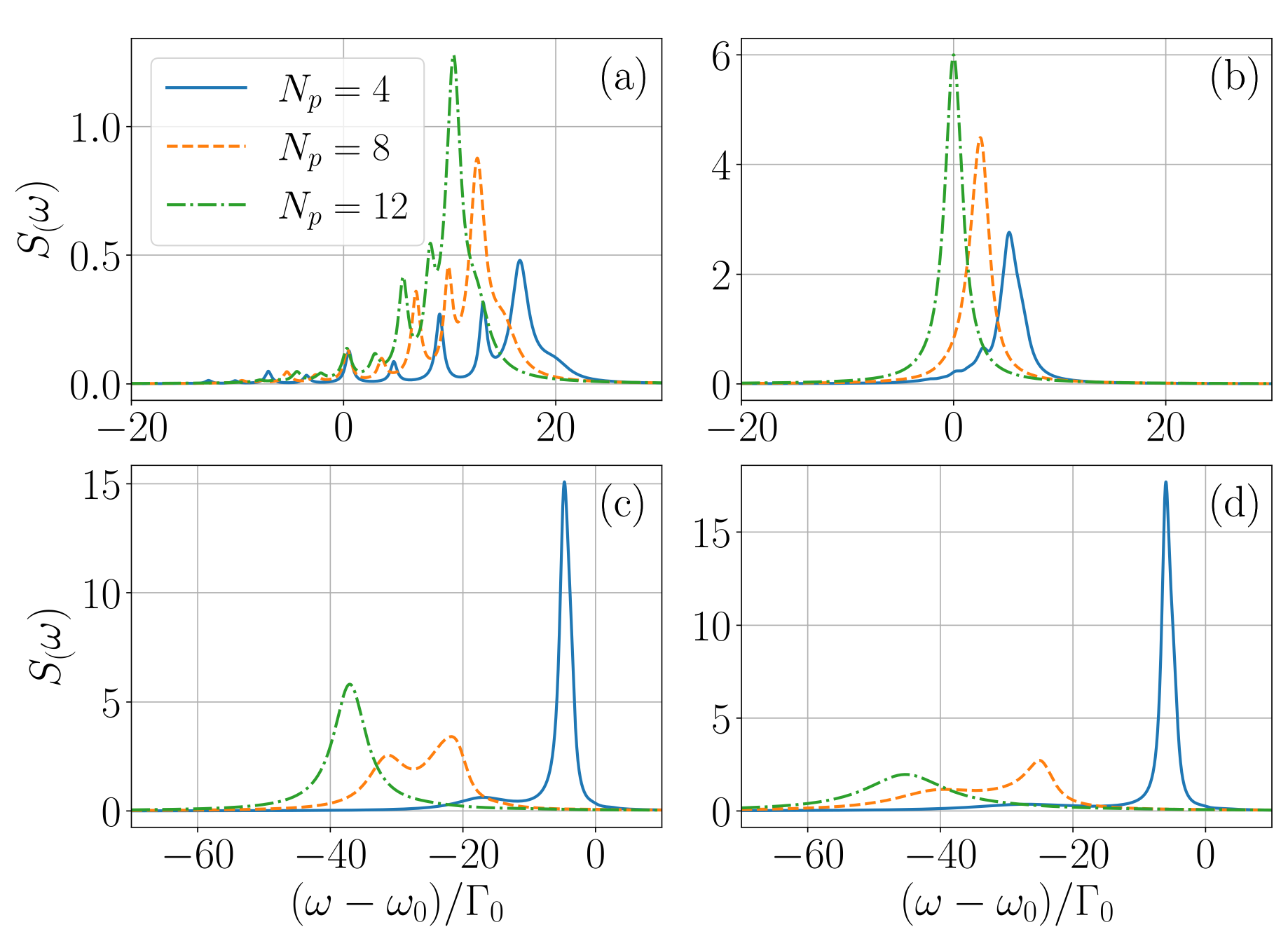}
    \caption{Steady-state light emission spectra in the direction of maximal emission for a chain of $N=12$ emitters spaced by $a = 0.05 \lambda_0$ for various pumped fractions $N_p = 4, 8, 12$ and pump powers (a)~$R=0.2\Gamma_0$, (b)~$R=1\Gamma$, (c)~$R=10\Gamma_0$, (d)~${R=20\Gamma_0}$. We observe a transition from multiple independent lines to synchronized emission around a single frequency. Note, that for a pumped fraction of less than half, a powerful narrow line emission persists even for a very strong pump. The area under the spectral curves corresponds to the total emitted power.
    }
    \label{fig:spectral_power}
\end{figure}

Here, we present the results for the steady-state total photon emission and spectral linewidth in the direction of maximal emission as a function of the total emitter number and a fixed pumped emitter number $N_p$. We obtain the direction of maximal emission $(\varphi,\theta)$ by calculating the electric field intensity distribution in the far field via Eq.~\eqref{e-field} and subsequently calculate the emission spectrum using Eq.~\eqref{spectrum}. The linewidth is defined as the full width at half maximum (FWHM) of the maximum peak in the spectrum, as the system can exhibit multiple peaks in any spatial direction, as shown in Fig.~\ref{fig:spectral_power}. 
Surprisingly, Fig.~\ref{fig3}\hyperref[fig3]{(b)} shows a sharp transition of the linewidth from $\Delta \nu\approx R+N_p\Gamma_0$ to $\Delta \nu\approx \Gamma_0$ as the total number of emitters $N$ is increased while keeping $N_p$ fixed. This is quite reminiscent of an ultracold cloud of atoms coupled to a single-mode cavity, where the system transitions to collective superradiant lasing above a certain pumping threshold~\cite{meiser2009prospects,bychek2021superradiant}. In the present case, the pumping rate $R=20\Gamma_0$ is chosen such that the system operates in the superradiant regime (colored as red in Fig.~\ref{fig:emission_rate}). As presented in Fig.~\ref{fig3}\hyperref[fig3]{(c)}, the spectral peak position approaches the atomic resonance frequency $\omega_0$ as the number of unpumped emitters $N-N_p$ increases, despite the initial strong dipole-dipole frequency shifts observed when all emitters are pumped.
\begin{figure}[t!]
    \centering
    \includegraphics[width=1\columnwidth]{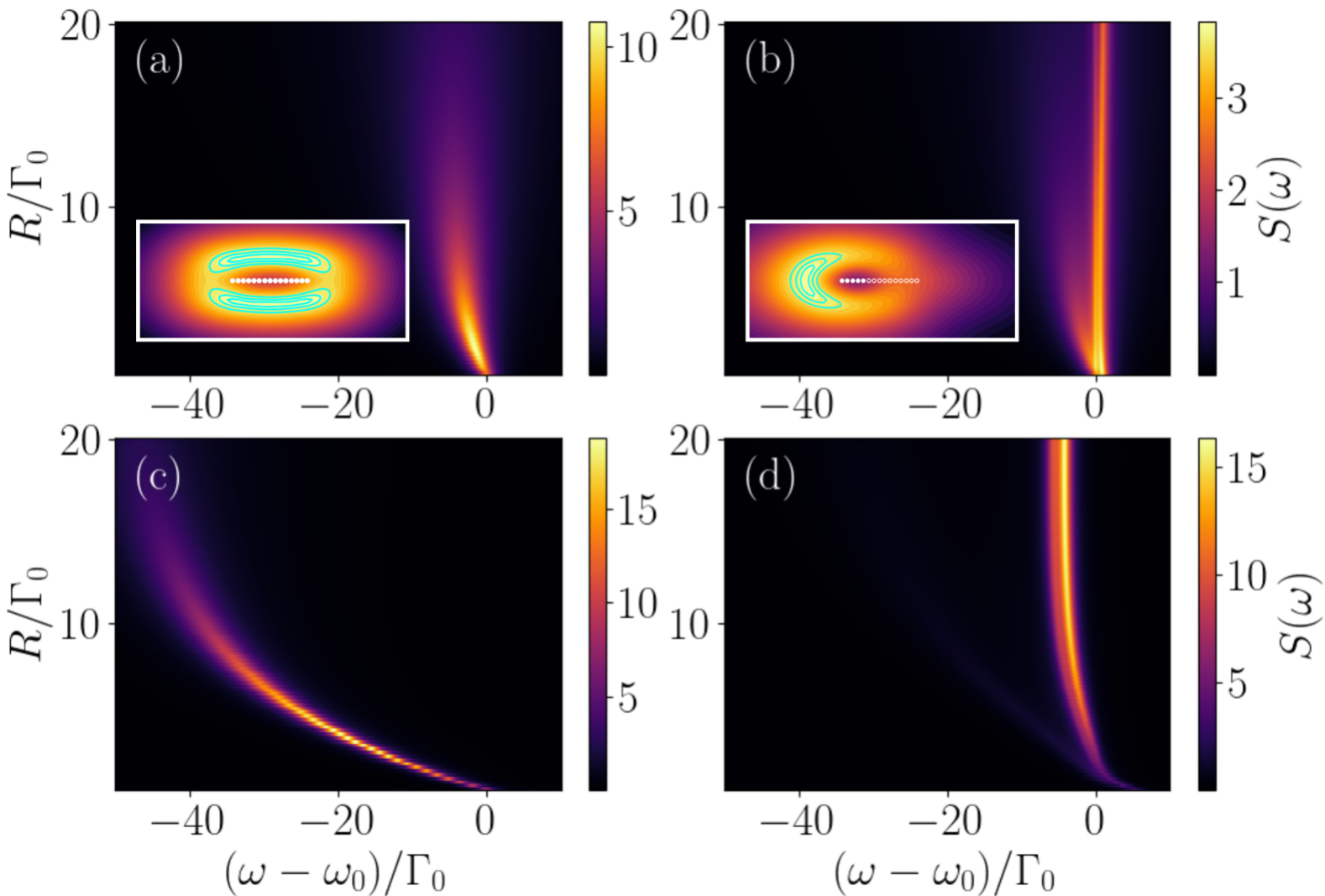}
    \caption{Pump power dependence of emission spectra in the direction of maximal emission for a chain of $N=15$ particles. The results are shown for emitter spacings of $a = 0.1\lambda_0$ with (a) $N_p/N = 1$ and (b) $N_p/N = 1/3$ pumped fraction, and for $a = 0.05\lambda_0$ with (c) $N_p/N = 1$ and (d) $N_p/N = 1/3$. The presence of non-pumped emitters leads to the appearance of a single narrow emission line close to the atomic resonance. Insets: The electric field intensity in the $xy$-plane for~${R=10\Gamma_0}$.}
    \label{fig6}
\end{figure}

\textit{Light emission directionality and spectra.}---Although the array's spatial extent considered thus far is only of the order of the optical wavelength, the emitted radiation pattern differs from the single-emitter dipolar pattern and exhibits directionality.
Simultaneously, the emitted light can exhibit a narrow spectral peak close to the bare emitter transition frequency $\omega_0$, if pumped above a certain threshold power, as shown in Fig.~\ref{fig:spectral_power} for $a=0.05\lambda_0$. This is of particular interest, as the system demonstrates superradiant emission with a narrow spectral linewidth without the need for optical elements or nearby resonators, unlike superradiant lasing in optical cavities~\cite{meiser2009prospects,kazakov2022ultimate,bychek2025self}. Furthermore, Fig.~\ref{fig6} shows the pump power dependence of the emission spectra for a partially (right) and fully pumped (left) array of $N=15$ emitters. In case of partial pumping, one can see a striking persistence of a single narrow spectral peak close to the bare transition frequency, which spectral intensity becomes particularly apparent at smaller spacings as can be seen in Fig.~\ref{fig6}\hyperref[fig6]{(d)}. The insets show the electric field intensity distribution in the $xy$-plane. 

\textit{Conclusions.}---Our results point toward a narrow-linewidth superradiant directional light source solely composed of a dense regular array of two-level dipole emitters under partial incoherent pumping. The collective light emission transitions from sub- to superradiance as the pump rate exceeds the individual spontaneous decay rate. While strong pumping provides population inversion, a particularly large emission gain is observed when the emitters are spaced at subwavelength distances, where the strong and directional narrow-band emission resembling lasing originates solely from resonant free-space dipole-dipole interaction.~The emission power and spectral purity, meaning a single spectral peak, are significantly enhanced when only a fraction of the atomic emitters are pumped and the remaining free atoms act as narrow-band out-coupling antennas.  

A laser-like action in the absence of an optical resonator has been previously experimentally observed in various systems and was termed mirrorless lasing~\cite{javan1961population,zhang2024extremely,akulshin2018continuous}. For a uniformly pumped ensemble, the superradiant power can reach twice the power of independent atomic emission. However, when additional unpumped emitters are present, strong pumping can lead to much larger gain factors only limited by the total number $N$ of unpumped and pumped atoms. The spectrum then splits into a single narrow peak close to the atomic resonance and a weak broad, strongly shifted peak directly from the pumped atoms. In the extreme case, only a small fraction of gain atoms can thus provide for strong narrow-band emission close to the bare atomic frequency, which is particularly favorable for an active optical frequency standard~\cite{ludlow2015optical,kazakov2022ultimate,norcia2018frequency}.

While we focused our model on the generic case of one-dimensional arrays, other geometries such as ring-shaped or planar arrays could prove to be effective~\cite{holzinger2020nanoscale}.
Geometric control over the atomic emitter ensemble can be expected to lead to improved spatial and spectral properties of the emitted light. We anticipate superior performance from much larger ensembles of emitters, which are currently out of reach of our numerical capabilities but potentially available in experiments.
To relax the requirements on tight atomic positioning, operations on narrow transitions in the (far-)infrared region could present a promising experimental approach.

\textit{Acknowledgments.}---The authors would like to thank D.\ Budker and S.\ F.\ Yelin for stimulating discussions. The numerical simulations were performed with the open-source frameworks QuantumCumulants.jl~\cite{plankensteiner2021quantumcumulantsjl} and  QuantumOptics.jl~\cite{kramer2018quantumoptics}. This work was supported by the FET OPEN Network Cryst3 funded by the European Union (EU) via Horizon 2020 and the FG5 from FWF. R.H. acknowledges funding by the NSF via PHY-2207972, the CUA PFC PHY-2317134.


\bibliographystyle{apsrev4-2}
\bibliography{references}

\newpage

\clearpage
\pagebreak
\onecolumngrid

\section*{Supplemental Material}
\setcounter{equation}{0}
\setcounter{figure}{0}
\setcounter{table}{0}
\makeatletter
\renewcommand{\fnum@figure}{FIG.~S\thefigure}
\makeatother
\renewcommand{\thesection}{S\arabic{section}} 
\renewcommand{\theequation}{S\arabic{equation}} 

Here we present the details on the second-order cumulant expansion approach, benchmarking the results against the master equation solution, and the calculation of the directional properties of light emission used in the main text. Next, we discuss the calculation of the second-order correlation function, as well as experimental considerations and the role of positional disorder.

\vspace{-0.45cm}

\section{Comparison of the cumulant expansion with the full quantum master equation solution} \label{appendix:comparison}

Using the full quantum master equation to describe the system's time evolution, means that the number of equations grows exponentially with the atom number, which limits the atom number that can be numerically simulated to small numbers. To study larger atom numbers, we apply the cumulant expansion method~\cite{kubo1962generalized,plankensteiner2021quantumcumulantsjl}, where one takes expectation values of the quantum Langevin equations of motion and truncates the set of equations by approximating averages of higher order operators with averages of lower order.
All atoms are assumed to be in the ground state initially, leading to $\langle \sigma^{ee}_n \rangle = 0$, $\langle \sigma^+_n \sigma^-_m \rangle=0$ and $\langle \sigma^{ee}_n \sigma^{ee}_m \rangle = 0$ at $t=0$ for all emitters $n,m$.

\begin{figure}[h!]
    \centering
    \includegraphics[width=0.73\columnwidth]{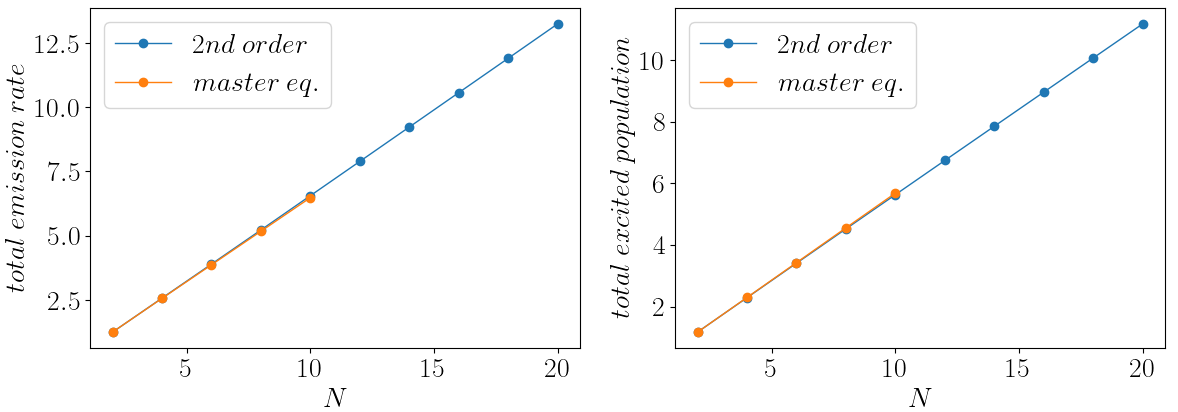}
    \caption{Comparison of the master equation solution (orange line) with the second-order cumulant expansion method (blue) for the total emission rate and excited state population plotted as a function of the total atom number $N$ in a fully pumped atomic ensemble. Parameters: $R = 1.5\Gamma_0, a = 0.1\lambda_0$.}
    \label{supp:fig1}
\end{figure}

\begin{figure}[h]
    \centering
    \includegraphics[width=\columnwidth]{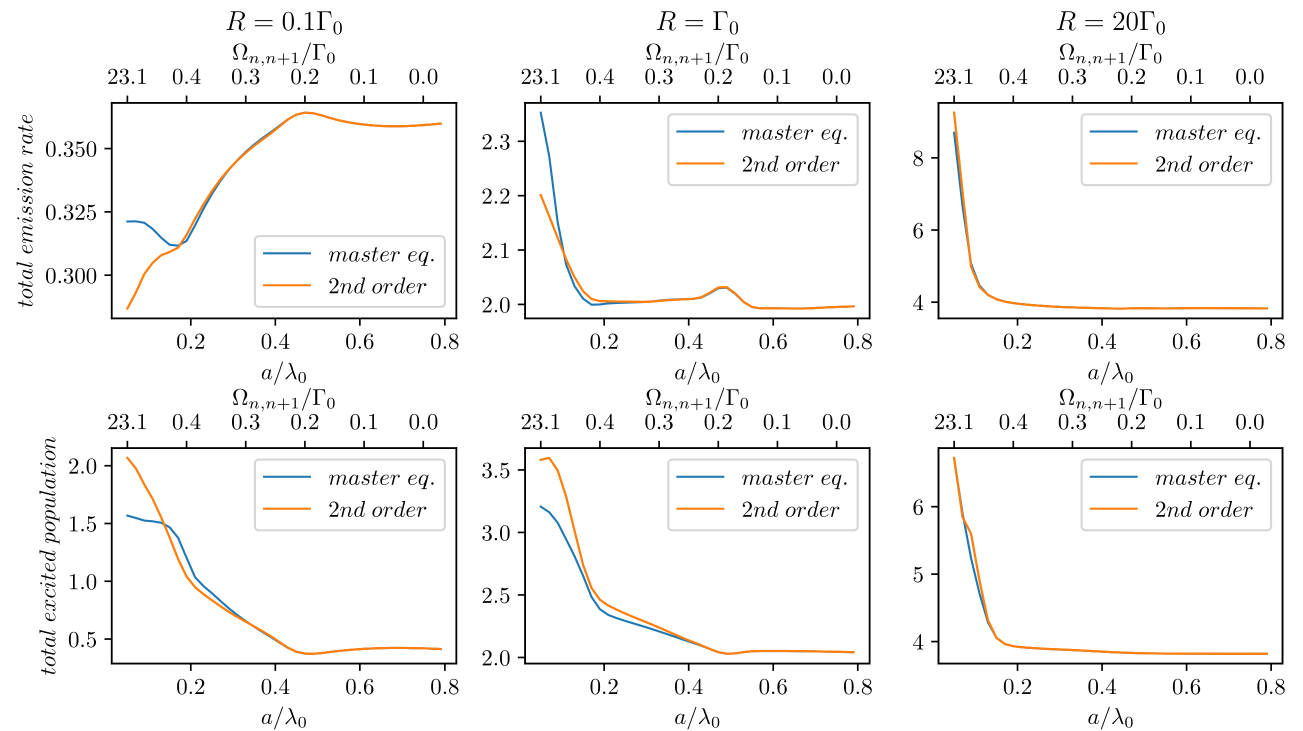}
    \caption{Comparison of the master equation solution (orange line) with the second-order cumulant expansion method (blue) for the total emission rate and excited state population plotted as a function of the chain parameter $a$ in a partially pumped atomic ensemble for three different pumping rates $R=0.1\Gamma_0$, $R=\Gamma_0$, and $R=20\Gamma_0$. Parameters: $N = 8, N_p=4$ and $\Omega_{n,n+1}$ denotes the coherent nearest-neighbor dipole-dipole coupling.}
        \label{supp:fig2}
\end{figure}

 We restrict to correlations up to the second order, where only three kinds of operators develop non-zero expectation values, namely $\langle \sigma^{ee}_n \rangle, \langle \sigma^+_n \sigma^-_m \rangle, \langle \sigma^{ee}_n \sigma^{ee}_m\rangle$ during the time dynamics. With this, we obtain a closed set of differential equations for the second-order cumulants
\begin{equation}
\label{eq.Heisenberg}
\begin{aligned}
\frac{d}{dt}  \langle \sigma^{ee}_n \rangle =& -(\Gamma_0+R_n) \langle \sigma^{ee}_n \rangle + 2\sum^N_{k \neq n} \Re \Big\{ g_{nk}\langle \sigma^+_k \sigma^-_n \rangle \Big\} +R_n\\
\frac{d}{dt}  \langle \sigma^+_n \sigma^-_m \rangle =& -\Big( \Gamma_0+\frac{R_n+R_m}{2}\Big) \langle \sigma^+_n \sigma^-_m \rangle + 2\Gamma_{nm}\langle \sigma^{ee}_n \sigma^{ee}_m\rangle + g_{nm}\langle \sigma^{ee}_m \rangle +g_{nm}^*\langle \sigma^{ee}_n \rangle \\
&- \sum_{k \neq n,m}^N \Big( g_{km}^*\langle \sigma^+_n \sigma^-_k \rangle \big(2\langle \sigma^{ee}_m \rangle -1 \big) + g_{kn} \langle \sigma^+_k \sigma^-_m \rangle \big(2\langle \sigma^{ee}_n \rangle -1 \big) \Big) \\
\frac{d}{dt}  \langle \sigma^{ee}_n \sigma^{ee}_m\rangle =& -(2\Gamma_0+R_n+R_m )\langle \sigma^{ee}_n \sigma^{ee}_m \rangle +R_n\langle \sigma^{ee}_m \rangle +R_m \langle \sigma^{ee}_n\rangle \\
&+2\sum^N_{k \neq n,m}\Re \Big\{ g_{km} \langle \sigma^{ee}_n \rangle \langle \sigma^+_k \sigma^-_m \rangle + g_{kn}\langle \sigma^{ee}_m\rangle \langle \sigma^+_k \sigma^-_n \rangle \Big\}, \\
\end{aligned}
\end{equation}
with $n\neq m$ and where we defined the collective dipole-dipole couplings $g_{nm} = i\Omega_{nm}- \Gamma_{nm}/2$. In Fig.~\hyperref[supp:fig1]{S}\ref{supp:fig1} and Fig.~\hyperref[supp:fig2]{S}\ref{supp:fig2} we show the comparison between the numerical solutions of Eq.~\eqref{eq.Heisenberg} and the quantum master equation calculating the total excited state population $\sum_n \langle \sigma^{ee}_n \rangle$ and the total emission rate $I(t) = \sum_{n,m} \Gamma_{nm}\langle \sigma^+_n\sigma^{-}_m \rangle $. We observe that at small interparticle spacings, the second-order cumulant solution deviates significantly from exact numerics for pumping rates $R \lesssim \Gamma_0$ (assuming identical pumping rates $R$), in particular for the excited state population. This is explained by the dominant role of subradiant collective states in the weak pump regime which demands higher-order operator correlations.
Overall, both the excited state population and emission rate show excellent agreement for $R>\Gamma_0$, as the emitter ensemble transitions from subradiance to steady-state superradiance (discussed in the main text).


\section{Directional properties of light emission: distribution of emission intensity and directional variation of spectra} \label{supp:spectrum} 

In order to analyze the emitted light in more detail, we compute its spectral linewidth. Therefore, we calculate the emission spectrum by means of the Wiener-Khinchin theorem. It is given as the Fourier transform of the first-order coherence function, $g^{(1)}(\tau) = \langle \mathbf{E}^-(\tau,\mathbf{r}) \mathbf{E}^+(0,\mathbf{r})\rangle $.~Typically, the spectrum exhibits a Lorentzian shape, which we characterize by the spectral linewidth $\Delta \nu$ as the full width at half maximum (FWHM).

\subsection*{Single atom with incoherent pump}
Firstly, we consider a single atom (or equivalently a collection of $N$ non-interacting atoms) with an incoherent pump rate $R$, since it can be solved analytically. To calculate the spectrum one needs to consider the time derivative of the first-order correlation $\langle \sigma^\dagger(\tau) \sigma(0)\rangle$. We do this by writing down the Quantum Langevin equation for the operator $\sigma$, which reads $\dot{\sigma} = -\frac{1}{2}(\Gamma_0+R) \sigma$ (in the rotating frame of the atomic frequency $\omega_0$) and where we assume the system initially to be in the vacuum state, thus neglecting input noise operators. This leads to the equation of motion for the first order correlation $\partial_\tau \langle \sigma^\dagger (\tau) \sigma (0)\rangle = -\frac{1}{2}(\Gamma_0+R) \langle \sigma^\dagger (\tau) \sigma (0)\rangle$, which can be easily integrated. Therefore, by using the Wiener-Khinchin theorem~\cite{puri2001mathematical} we find that the spectrum for non-interacting atoms is given by
\begin{align}
    S(\omega)_\mathrm{single} &= 2 \mu_0^2 \omega_0^4 \ \Re \bigg\{  \int_0^\infty d \tau e^{-i \omega \tau} \langle \sigma^+(\tau) \sigma^-(0)\rangle \bigg\} \nonumber \\
    &=2  \mu_0^2 \omega_0^4 \ \langle \sigma^+ \sigma^-\rangle_\mathrm{st} \ \Re \bigg\{  \int_0^\infty d \tau  e^{-i \omega \tau} e^{-(R+\Gamma_0) \tau/2} \bigg\}  =  \frac{2 R\Gamma_0}{4\omega^2+(\Gamma_0+R)^2} ,
\end{align}
where we have used the steady state value $\langle \sigma^+ \sigma^-\rangle_\mathrm{st} = R/(R+\Gamma_0)$.
This spectrum exhibits a linewidth (FWHM) $\Delta\nu = \Gamma_0 + R$. The spectrum is not directional, has a maximum value $2R\Gamma_0/(\Gamma_0+R)^2$ at $\omega=0$, and the linewidth broadens with increasing pumping strength $R$.
\subsection*{Ensemble of \textit{N} interacting atoms}

\begin{figure}[b!]
    \centering
\includegraphics[width=0.5\columnwidth]{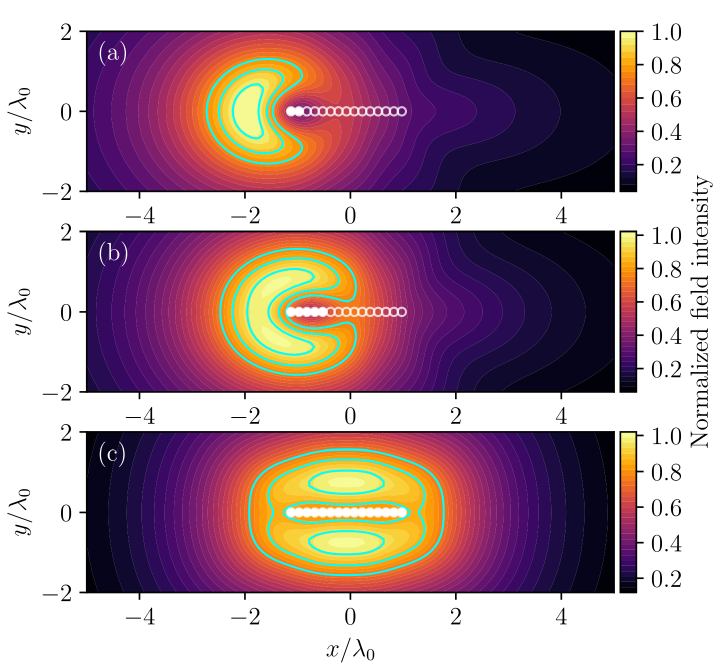}
    \caption{Directional emission under partial pumping: Steady-state electric field intensity distribution calculated using a second-order cumulant expansion method. The normalized intensity is plotted in the $xy$-plane seen from the $z$-direction at $z=\lambda_0$ for the case of (a)~$N_p=2$, (b) $N_p=5$, (c) $N_p=15$ pumped emitters, where $N=15$, $R=10\Gamma_0$, $a = 0.15\lambda_0$.}
    \label{fig:directional_emission}
\end{figure}

As the total size of emitting structures can go beyond the emitted light wavelength, we also expect a modification of the angular emission pattern compared to a single emitter. In Fig.~\hyperref[fig:directional_emission]{S}\ref{fig:directional_emission} we calculate the steady-state electric field intensity $\langle \mathbf{E}^+(\mathbf{r}) \mathbf{E}^-(\mathbf{r})\rangle$ in the $xy$-plane for a (partially) pumped $N=15$ emitter chain with the spacing of $a=0.15\lambda_0$, leading to directional light emission along the chain axis under partial pumping. Here, where we have used the expression for the electric field operator $\mathbf{E}^-(\mathbf{r}) = \mu_0\omega_0^2 \sum_{n=1}^N \mathbf{G}(\mathbf{r-r_n})\cdot \mathbf{d}_n \sigma_n^-$ introduced in the main text. The collective emission exhibits directionality at smaller pumped fractions $N_p/N<1$ while for $N_p/N=1$ the emission resembles a single dipole radiation pattern with superradiant emission perpendicular to the chain direction.

\begin{figure}[h!]
    \centering
    \includegraphics[width=0.55\columnwidth]{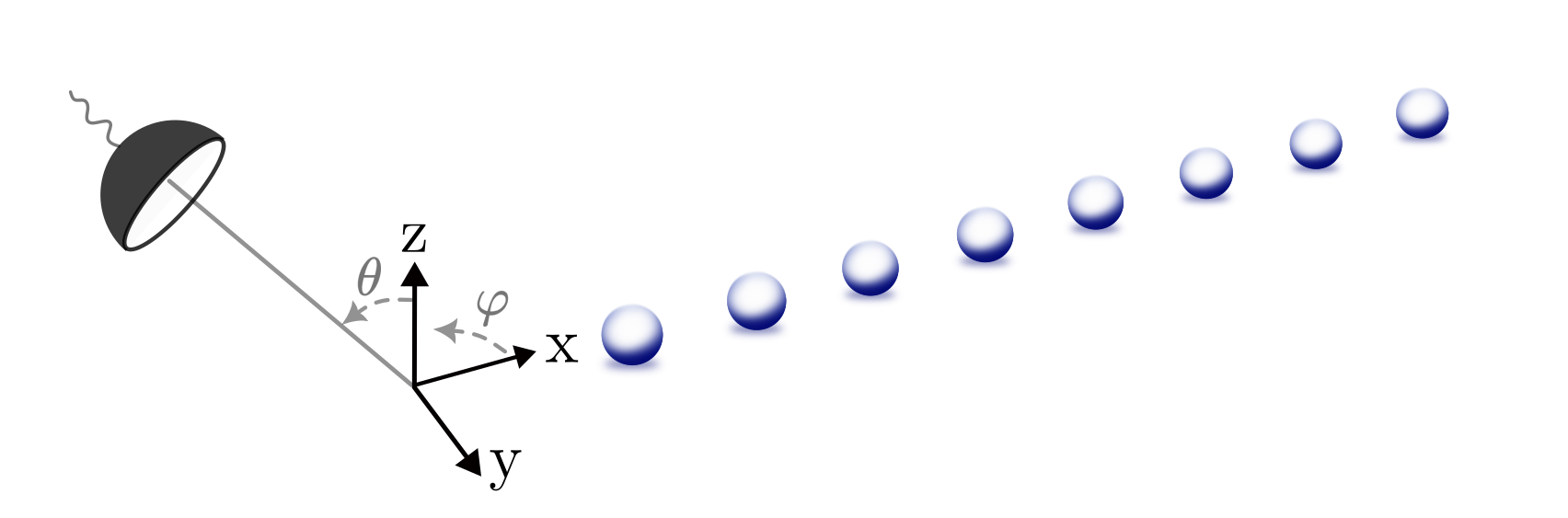}
    \caption{Directional light emission from a chain of atoms. The atomic emitters are positioned along the $x$ direction, with photon detection in the far field ($r \gg \lambda_0$). The detection point occurs in the direction with angles $(\varphi,\theta)$ as shown in the schematic. For instance, detection in the $x$ direction corresponds to $(\varphi,\theta)=(0,\pi/2)$.}
    \label{supp:fig-chain}
\end{figure}

The directional emission spectrum for $N$ dipole-coupled atomic two-level emitters is calculated as the Fourier transformation of the first order correlation $\langle \mathbf{E}^+(\tau,\mathbf{r}) \mathbf{E}^-(0,\mathbf{r})\rangle$ and reads
\begin{equation}
\label{Spectrum}
\begin{aligned}
S(\omega, \mathbf{r}) &= 2 \Re \left\{\int_{0}^{\infty} d\tau e^{-i\omega \tau} \langle \mathbf{E}^+(\tau, \mathbf{r}) \mathbf{E}^-(0, \mathbf{r}) \rangle \right\}\\
 &= 2\mu_0^2 \omega_0^4 \Re \left\{ \int_{0}^{\infty} d\tau e^{-i\omega \tau} \sum_{n,m=1}^N \langle \sigma^+_n(\tau) \sigma^-_m(0) \rangle \; \mathbf{d}_m^*  \cdot \mathbf{G}^*(\mathbf{r}-\mathbf{r}_m) \cdot \mathbf{G}(\mathbf{r}-\mathbf{r}_n) \cdot \mathbf{d}_n \right\}.
\end{aligned}
\end{equation}
The time evolution of the expectation values $\langle \sigma^+_n(\tau) \sigma^-_m(0) \rangle$ can be obtained using the second-order cumulant expansion and the quantum regression theorem,
\begin{equation}
\label{Correlation_function_spectrum}
\frac{d}{d\tau}  \sum_{m=1}^N \langle \sigma^+_n (\tau) \sigma^-_m(0) \rangle = \sum_{k \neq n}^N \Big((i\Omega_{kn}-\Gamma_{kn}/2)(1-2\langle \sigma^+_n \sigma^-_n \rangle_\mathrm{st})\sum_{m=1}^N \langle \sigma^+_k (\tau) \sigma^-_m(0) \rangle \Big)  - \frac{1}{2}(\Gamma_0+R_n) \sum_{m=1}^N \langle \sigma^+_n (\tau) \sigma^-_m(0) \rangle,
\end{equation}
which can be integrated and allows to evaluate the spectrum at any point $\mathbf{r}$ in space. Next, we are interested in the far field values of the spectrum and consider the limit $|\mathbf{r}|\gg \lambda_0$
in which case it simplifies to~\cite{carmichael2000quantum,masson2020many}
\begin{equation}
\label{Correlation_function_spectrum_far_field}
S(\omega, \varphi,\theta) = \frac{3\Gamma_0}{8\pi} \Big(1- (\mathbf{R}(\theta,\varphi)\cdot \mathbf{d})^2\Big) \times 2 \Re \left\{ \int_{0}^{\infty} d\tau e^{-i\omega \tau} \sum_{n,m=1}^N  e^{i k_0 (x_m - x_n)\cos{\varphi}\sin{\theta}} \langle  \sigma^+_n(\tau) \sigma^-_m(0) \rangle \right\}.
\end{equation}
 Here, we assume a one-dimensional ensemble of emitters positioned along the $x$ direction with identical dipole polarizations $\mathbf{d}$, total chain length $L=aN$. The direction of detection is quantified by the unit vector $\mathbf{R}(\theta,\varphi)=[\cos(\varphi)\sin(\theta),\sin(\varphi)\sin(\theta),\cos(\theta) ]$ illustrated in Fig.~\hyperref[supp:fig-chain]{S}\ref{supp:fig-chain}. For instance, by choosing linear polarization in $z$-direction as in the main text, the scalar product $\mathbf{R}(\theta,\varphi)\cdot \mathbf{d}$ vanishes for detection in the $xy$-plane.
In the main text we show how the directional spectrum can exhibit a spectral linewidth $\Delta \nu$ smaller or comparable with $\Gamma_0$ even for steady-state superradiant emission. This stands in stark contrast to the expected linewidth of free-space superradiance with $\Delta \nu \sim N\Gamma_0$ or the case of non-interacting emitters with $\Delta \nu \sim \Gamma_0+R$.

\section{Second-order correlation function} \label{supp:g2} 
\begin{figure}[ht!]
    \centering
\includegraphics[width=1\columnwidth]{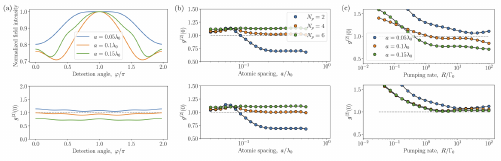}
    \caption{An equidistant chain of $N=8$ atoms along the $x$-direction, each being polarized linearly along the $z$-direction. (a) The normalized steady-state field intensity and normalized second-order correlation function $g^{(2)}(0)$ in the $xy$-plane as a function of the azimuthal angle $\varphi$. The incoherent pumping is $R=20\Gamma_0$ and $N_p=2$ (The first two most left atoms are pumped). (b) The second-order correlation in direction of maximal emission $(\varphi=\pi)$ as a function of atomic spacing for $R=20\Gamma_0$ (top) and $R=40\Gamma_0$ (bottom). (c) The second-order correlation in direction of maximal emission for $N_p=2$ (top) and $N_p=4$ (bottom) as a function of the pumping rate. }
    \label{fig:supp-g2}
\end{figure}
In order to study the statistical properties of the emitted light we calculate the normalized second-order correlation at zero time delay of the emitted electric field. Similar to the electric field intensity $\langle \mathbf{E}^+(\mathbf{r}) \mathbf{E}^-(\mathbf{r}) \rangle$ above we evaluate the normalized second-order correlation functions (with zero-time delay, $\tau=0$)
\begin{equation}
    g^{(2)}(0) = \frac{ \big\langle (\mathbf{E}^+(\mathbf{r}))^2 (\mathbf{E}^-(\mathbf{r}))^2 \big\rangle}{\langle \mathbf{E}^+(\mathbf{r}) \mathbf{E}^-(\mathbf{r}) \rangle^2 },
\end{equation}
using the electric field operators at position $\mathbf{r}$. We will evaluate $g^{(2)}(0)$ in the far field ($|\mathbf{r}|= r \gg \lambda_0$) and as a function of the emission azimuthal angle $\varphi$, defined in the plane containing the array and in the direction
of maximal emission, illustrated in Fig.~\hyperref[supp:fig-chain]{S}\ref{supp:fig-chain}. Coherent light follows a Poissonian photon statistics which implies $g^{(2)}(0)$ close to unity, whereas $g^{(2)}(0)$ significantly bigger than one is indicative of amplified spontaneous emission (bunched light). In Fig.~\hyperref[fig:supp-g2]{S}\ref{fig:supp-g2}, we present the second-order correlation function for an equidistant chain of $N=8$ atoms and observe that its steady-state values stay close to unity above the pumping threshold, in particular at spacings $a\lesssim 0.1\lambda_0$, where the emission is superradiant.

\section{Experimental considerations}

Experimental realization of mirrorless superradiant lasing in subwavelength-spaced emitter arrays hinges on engineering dipole–dipole couplings strong enough to enforce collective emission, which in practice requires mean separations $a\!\lesssim\!\lambda_0/4$. Although challenging, this regime might be reached with existing and rapidly progressing neutral atom platforms. To relax the requirements on tight atomic positioning, operations on narrow transitions in the (far-)infrared region could present a promising experimental approach. 
Subwavelength optical lattices based on short-wavelength transitions provide a feasible route that supports large ensembles with alkaline-earth atoms~\cite{Du2024_dipolar_bilayer,Baier2016,Rui2020}.
Alternatively, optical tweezer arrays employ positioning techniques which can arrange tens or even thousands of atoms in arrays with high accuracy and with single-site reconfigurability~\cite{Seubert2025, Norcia2018,Pause2024laser}. Using highly-excited Rydberg states, resonant dipole-dipole interactions in the microwave regime can be leveraged to facilitate strong coupling between the emitters~\cite{brow2017optical,glicenstein2020collective,barredo2015coherent}.

Furthermore, solid-state emitters could offer unique advantages in emitter positioning, although substantial inhomogeneous broadening of the individual emitter resonance frequencies remains a considerable challenge.
In Ref.~\cite{bychek2021superradiant}, we showed that under strong pumping the synchronization of atomic dipoles with inhomogeneous frequency broadening can be observed, provided the pumping strength exceeds the standard deviation of the broadening. 
Site-controlled quantum-dot platforms now achieve dense self-organized ordered arrays with continuous progress toward minimizing inhomogeneous broadening~\cite{huggenberger2011narrow, Grosse2020}.
These platforms naturally support incoherent electrical or optical pumping and benefit from scalability of the emitter number.

\begin{figure}[ht!]
    \centering
\includegraphics[width=0.95\columnwidth]{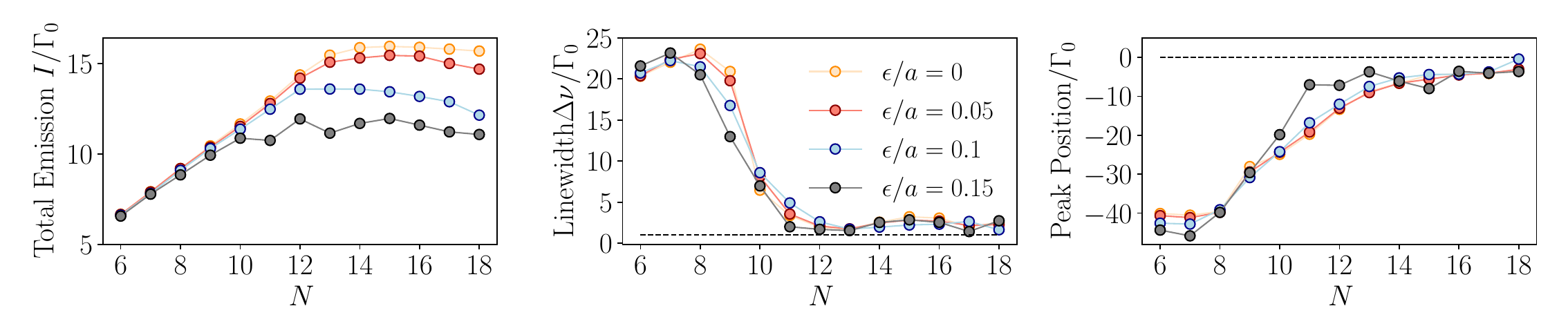}
    \caption{Photon emission, spectral linewidth and spectral peak position for $N_p=6$ pumped atoms as a function of atom number. The positional disorder is modeled by random fluctuations of the initially ordered atomic positions (along $x$) in the $x$- and $y$-directions with standard deviation $\epsilon/a$. Each curve is an average over 50 realizations. Pumping rate $R=20\Gamma_0$ and initial chain spacing $a=0.05\lambda_0$.}
    \label{fig:disorder}
\end{figure}

Most experiments including optical tweezers and quantum-dot arrays show fluctuations in the emitter positions deviating from an ideal array implementation on the order of $5\%-10 \%$ or more~\cite{hsu2022single,holman2024trappingsingleatomsmetasurface}. In Fig.~\hyperref[fig:disorder]{S}\ref{fig:disorder} we show the effect of positional disorder in a linear chain of $N$ atoms placed along the $x$-direction. The atomic positions fluctuate in the $xy$-plane and one can observe that the results of a perfectly ordered chain remain valid under experimentally realistic values. Even though the steady-state photon emission is decreasing with increasing disorder, the linewidth and spectral peak position remain essentially unaffected.

\end{document}